\documentclass{article}
\usepackage{spconf,amsmath,graphicx}

\usepackage{amsmath,amssymb,epsfig,psfrag,cite,subfigure}
\include{macros}
\usepackage{graphicx}
\usepackage{epstopdf}

\usepackage{bm}

\usepackage{tcolorbox}

\usepackage{soul}

\usepackage{amsthm}

\usepackage{nicefrac}

\usepackage{lipsum,multicol}

\usepackage{algorithm}
\usepackage{algpseudocode}

\usepackage{enumitem}
\usepackage{mwe}    
\usepackage{epsfig,psfrag}
\usepackage{subfigure}
\usepackage{color}
\usepackage{url}
\usepackage{mathtools,xparse}

\usepackage{gensymb}

\usepackage[multiple]{footmisc}

\theoremstyle{remark}

\newtheoremstyle{mytheoremstyle} 
    {\topsep}                    
    {\topsep}                    
    {\upshape}                   
    {.5em}                           
    {\itshape}                   
    {.}                          
    {.5em}                       
    {}  
    
\theoremstyle{mytheoremstyle}

\newtheoremstyle{iremark}
  {\topsep}   
  {\topsep}   
  {\upshape}  
  {0.2in}       
  {\itshape}  
  {.}         
  {5pt plus 1pt minus 1pt} 
  {\thmname{#1}\thmnumber{ \itshape#2}\thmnote{ (#3)}} 

\theoremstyle{iremark}

\newtheorem{remark}{Remark}

\newtheorem*{proof}{Proof}

\DeclarePairedDelimiter\abs{\lvert}{\rvert}%

\makeatletter \renewcommand\d[1]{\ensuremath{%
		\;\mathrm{d}#1\@ifnextchar\d{\!}{}}}
\makeatother

\makeatletter
\newcommand*\rel@kern[1]{\kern#1\dimexpr\macc@kerna}
\newcommand*\widebar[1]{%
  \begingroup
  \def\mathaccent##1##2{%
    \rel@kern{0.8}%
    \overline{\rel@kern{-0.8}\macc@nucleus\rel@kern{0.2}}%
    \rel@kern{-0.2}%
  }%
  \macc@depth\@ne
  \let\math@bgroup\@empty \let\math@egroup\macc@set@skewchar
  \mathsurround\z@ \frozen@everymath{\mathgroup\macc@group\relax}%
  \macc@set@skewchar\relax
  \let\mathaccentV\macc@nested@a
  \macc@nested@a\relax111{#1}%
  \endgroup
}
\makeatother

\newcommand{\norm}[1]{\left\lVert#1\right\rVert}

\newcommand{\normbig}[1]{\Big\lVert#1\Big\rVert}

\newcommand{\thnew}[1]{ {#1^{\rm{th} } } }

\newcommand{\Tsym}{ T_{\rm{sym}} }
\newcommand{\Tcp}{ T_{\rm{cp}} }

\newcommand{\FF}{ \mathbf{F} }

\newcommand{\deltaf}{ \Delta f }
\newcommand{\fc}{ f_c }
\newcommand{\Ntx}{ N_{\rm{T}} }
\newcommand{\Nrx}{ N_{\rm{R}} }
\newcommand{\ftx}{ \ff_{\rm{T}} }

\newcommand{\atx}{ \aaa_{\rm{T}} }
\newcommand{\arx}{ \aaa_{\rm{R}} }

\newcommand{\wwconj}{ \ww^{*} }
\newcommand{\wwhat}{ \widehat{\ww} }

\newcommand{\sm}{ s_{m} }
\newcommand{\hhhat}{ \widehat{\hh} }

\newcommand{\projrange}[1]{\boldsymbol{\Pi}_{#1}}

\newcommand{\xnm}{ x_{n,m} }

\newcommand{\boldHhat}{ \widehat{\boldH} }

\newcommand{\alphabarhat}{ \widehat{\alphabar} }

\newcommand{\rect}[1]{ { \rm{rect} }\left(#1\right) }

\newcommand{\Smcal}{ \mathcal{S} }

\newcommand{\mtCN}{{\mathcal{CN}}}

\newcommand{\vecc}[1]{ {\rm{vec}}\left(#1\right)  }

\newcommand{\diag}[1]{ {\rm{diag}}\left(#1\right)  }

\newcommand{\Imatrix}{{ \boldsymbol{\mathrm{I}} }}

\newcommand{\aaa}{\mathbf{a}}
\newcommand{\cc}{ \mathbf{c} }
\newcommand{\bb}{ \mathbf{b} }

\newcommand{\nuhat}{{ \widehat{\nu} }}
\newcommand{\tauhat}{{ \widehat{\tau} }}
\newcommand{\thetahat}{{ \widehat{\theta} }}

\newcommand{\boldzero}{{ {\boldsymbol{0}} }}

\newcommand{\ff}{\mathbf{f}}

\newcommand{\boldY}{ \mathbf{Y} }

\newcommand{\boldX}{ \mathbf{X} }

\newcommand{\Xbbar}{ \widebar{\boldX} }
\newcommand{\Ybbar}{ \widebar{\boldY} }
\newcommand{\Zbbar}{ \widebar{\boldZ} }

\newcommand{\boldZ}{ \mathbf{Z} }

\newcommand{\boldQ}{ \mathbf{Q} }
\newcommand{\boldH}{ \mathbf{H} }

\newcommand{\boldR}{ \mathbf{R} }

\newcommand{\boldU}{ \mathbf{U} }

\newcommand{\boldLambda}{ \mathbf{\Lambda} }

\newcommand{\yy}{ \mathbf{y} }

\newcommand{\xx}{ \mathbf{x} }
\newcommand{\hh}{ \mathbf{h} }
\newcommand{\ww}{ \mathbf{w} }

\newcommand{\DD}{ \mathbf{D} }

\newcommand{\boldSigma}{ \mathbf{\Sigma} }

\newcommand{\transpose}[1]{ {#1}^{T} }

\newcommand{\complexset}[2]{ \mathbb{C}^{#1 \times #2}  }

\newcommand{\alphabar}{ \widebar{\alpha} }

\title{ICI-Aware Parameter Estimation for MIMO-OFDM Radar via APES Spatial Filtering}
\name{Musa Furkan Keskin\textsuperscript{*}, Henk Wymeersch\textsuperscript{*}, Visa Koivunen\textsuperscript{$\dagger$}
\thanks{This work is supported by Vinnova grant 2018-01929 and the Marie Sk\l{}odowska-Curie Individual Fellowships (H2020-MSCA-IF-2019) Grant 888913 (OTFS-RADCOM).} 
}
\address{\textsuperscript{*}Chalmers University of Technology, Gothenburg, Sweden \\ \textsuperscript{$\dagger$}Aalto University, Espoo, Finland}
\begin{document}
\ninept
\maketitle
\begin{abstract}\vspace{-0.05in}
We propose a novel three-stage delay-Doppler-angle estimation algorithm for a MIMO-OFDM radar in the presence of inter-carrier interference (ICI). First, leveraging the observation that spatial covariance matrix is independent of target delays and Dopplers, we perform angle estimation via the MUSIC algorithm. For each estimated angle, we next formulate the radar delay-Doppler estimation as a joint carrier frequency offset (CFO) and channel estimation problem via an APES (amplitude and phase estimation) spatial filtering approach by transforming the delay-Doppler parameterized radar channel into an unstructured form. In the final stage, delay and Doppler of each target can be recovered from target-specific channel estimates over time and frequency. Simulation results illustrate the superior performance of the proposed algorithm in high-mobility scenarios. 
\end{abstract}
\vspace{-0.1in}
\section{Introduction}
\label{sec_intro}
\vspace{-0.1in}
With a huge increase of interest in co-existence and spectrum sharing among radar and communications, joint radar-communications (JRC) strategies are being actively developed for 5G and beyond wireless systems \cite{jointRadCom_review_TCOM,SPM_JRC_2019,SPM_Zheng_2019,chiriyath2017radar,Eldar_SPM_JRC_2020,Canan_SPM_2020}. A promising approach to practical JRC implementation is to design dual-functional radar-communications (DFRC) systems, which employ a single hardware that can simultaneously perform radar sensing and data transmission with a co-designed waveform \cite{DFRC_SPM_2019,DFRC_Waveform_Design,SPM_JRC_2019}. Orthogonal
frequency-division multiplexing (OFDM) has been widely investigated as a DFRC waveform due its wide availability in wireless communication systems and its potential to achieve high radar performance \cite{RadCom_Proc_IEEE_2011,General_Multicarrier_Radar_TSP_2016,ICI_OFDM_TSP_2020}. In the literature, estimator design for OFDM radar sensing has been studied in both single-antenna \cite{RadCom_Proc_IEEE_2011,Firat_OFDM_2012,ICI_OFDM_TSP_2020} and multiple-input multiple-output (MIMO) \cite{mmWave_JRC_TAES_2019,MIMO_OFDM_radar_TAES_2020} settings.

In high-mobility scenarios, such as millimeter-wave (mmWave) vehicular JRC systems \cite{SPM_JRC_2019}, Doppler-induced intercarrier interference (ICI) can significantly degrade the performance of OFDM from both radar and communications perspective \cite{OFDM_ICI_TVT_2017,ICI_OFDM_radar_2015,multiCFO_TWC_2018}. To improve OFDM radar performance, various ICI mitigation approaches have been proposed  \cite{Firat_OFDM_2012,OFDM_ICI_TVT_2017,ICI_OFDM_radar_2015,ICI_OFDM_TSP_2020}. In \cite{OFDM_ICI_TVT_2017}, the ICI effect is eliminated via an all-cell Doppler correction (ACDC) method, which requires the OFDM symbol matrix to be rank-one and therefore leads to a substantial loss in data rate. Similarly, ICI mitigation technique in \cite{ICI_OFDM_radar_2015} imposes certain constraints on transmit symbols, which impedes dual-functional operation. An alternating maximization approach is designed in \cite{ICI_OFDM_TSP_2020} to reduce the complexity of high-dimensional maximum-likelihood (ML) search by assuming that the number of targets is known a-priori. The work in \cite{Firat_OFDM_2012} considers a single-target scenario and proposes a pulse compression technique to compensate for ICI-induced phase rotations across OFDM subcarriers.


In this paper, we propose an ICI-aware delay-Doppler-angle estimation algorithm for a MIMO-OFDM radar with arbitrary transmit symbols in a generic multi-target scenario. 
The main ingredient is to re-formulate radar sensing as a joint carrier frequency offset (CFO)\footnote{Borrowing from the OFDM communications literature \cite{multiCFO_TWC_2018}, we use Doppler and CFO interchangeably throughout the text.} and channel estimation problem, which allows us to decontaminate the ICI effect from the resulting channel estimates, leading to high-accuracy delay-Doppler estimation.
To that end, we first perform angle estimation using the MUSIC high-resolution direction finding algorithm \cite{MUSIC_1986} based on the observation that spatial covariance matrix does not depend on target delays and Dopplers. To suppress mutual multi-target interferences \cite{Est_MIMO_radar_2013} in the spatial domain, we then devise an APES-like spatial filtering approach \cite{Est_MIMO_radar_2008,jointRadCom_review_TCOM} that performs joint Doppler/CFO and unstructured radar channel estimation for each estimated target angle separately. Finally, delay-Doppler of each target can be estimated from target-specific channel estimates by exploiting the OFDM time-frequency structure. Simulations are carried out to demonstrate the performance of the proposed algorithm in high-mobility scenarios. To the best of authors' knowledge, this is the first algorithm for MIMO-OFDM radar that takes into account the effect of ICI in estimating multiple target parameters without imposing any structure on data symbols. 

\vspace{-0.03in}
\textit{Notations:}  Uppercase (lowercase) boldface letters are used to denote matrices (vectors). $(\cdot)^{*}$, $(\cdot)^{T}$ and $(\cdot)^{H}$ represent conjugate, transpose and Hermitian transpose operators, respectively. $\Re \left\{ \cdot \right \}$ denotes the real part. The $\thnew{n}$ entry of a vector $\xx$ is denoted as $\left[\xx\right]_i$, while the $\thnew{(m,n)}$ element of a matrix $\boldX$ is $\left[ \boldX \right]_{m,n}$. $\projrange{\boldX} = \boldX (\boldX^H \boldX)^{-1} \boldX^H$ represents the orthogonal projection operator onto the column space of $\boldX$ and $\odot$ denotes the Hadamard product.

\vspace{-0.2in}
\section{OFDM Radar System Model}
\vspace{-0.1in}
Consider an OFDM DFRC transceiver that communicates with an OFDM receiver while concurrently performing radar sensing using the backscattered signals for target detection \cite{RadCom_Proc_IEEE_2011,DFRC_SPM_2019}. The DFRC transceiver is equipped with an $\Ntx$-element transmit (TX) uniform linear array (ULA) and an $\Nrx$-element receive (RX) ULA.
We assume co-located and perfectly decoupled TX/RX antenna arrays so that the radar receiver does not suffer from self-interference due to full-duplex radar operation \cite{Interference_MIMO_OFDM_Radar_2018,RadCom_Proc_IEEE_2011,OFDM_Radar_Phd_2014,80211_Radar_TVT_2018}. In this section, we derive OFDM transmit and receive signal models and formulate the multi-target parameter estimation problem. 



\vspace{-0.1in}
\subsection{Transmit Signal Model}\label{sec_transmit}
\vspace{-0.1in}
We consider an OFDM communication frame consisting of $M$ OFDM symbols, each of which has a total duration of $\Tsym = \Tcp + T$ and a total bandwidth of $N \deltaf = B$. Here, $\Tcp$ and $T$ denote, respectively, the cyclic prefix (CP) duration and the elementary symbol duration, $\deltaf = 1/T$ is the subcarrier spacing, and $N$ is the number of subcarriers \cite{RadCom_Proc_IEEE_2011}. Then, the complex baseband transmit signal for the $\thnew{m}$ symbol is given by
\begin{equation}\label{eq_ofdm_baseband}
\sm(t) = \frac{1}{\sqrt{N}} \sum_{n = 0}^{N-1}  \xnm \, e^{j 2 \pi n \deltaf t} \rect{\frac{t - m\Tsym}{\Tsym}}
\end{equation} 
where $\xnm$ denotes the complex data symbol on the $\thnew{n}$ subcarrier for the $\thnew{m}$ symbol \cite{General_Multicarrier_Radar_TSP_2016}, and $\rect{t}$ is a rectangular function that takes the value $1$ for $t \in \left[0, 1 \right]$ and $0$ otherwise. Assuming a single-stream beamforming model \cite{80211_Radar_TVT_2018,mmWave_JRC_TAES_2019,MIMO_OFDM_Single_Stream}, the transmitted signal over the block of $M$ symbols for $t \in \left[0, M \Tsym \right]$ can be written as 
\begin{equation}\label{eq_passband_st}
\Re \left\{ \ftx \sum_{m = 0}^{M-1} \sm(t) e^{j 2 \pi \fc t} \right\}
\end{equation}
where $\fc$ and $\ftx \in \complexset{\Ntx}{1}$ denote, respectively, the carrier frequency and the TX beamforming vector.
\vspace{-0.1in}

\subsection{Receive Signal Model}\label{sec_radar_rec}
Suppose there exists a point target in the far-field, characterized by a complex channel gain $\alpha$ (including path loss and radar cross section effects), an azimuth angle $\theta$, a round-trip delay $\tau$ and a normalized Doppler shift $\nu = 2 v/c$ (leading to a time-varying delay $\tau(t) = \tau - \nu t$), where $v$ and $c$ denote the radial velocity and speed of propagation, respectively.
In addition, let $\atx(\theta) \in \complexset{\Ntx}{1}$ and $\arx(\theta) \in \complexset{\Nrx}{1}$ denote, respectively, the steering vectors of the TX and RX ULAs, i.e., $\left[ \atx(\theta) \right]_i = e^{j \frac{2 \pi}{\lambda} d i \sin(\theta)}$ and $\left[ \arx(\theta) \right]_i = e^{j \frac{2 \pi}{\lambda} d i \sin(\theta)}$,
where $\lambda$ and $d = \lambda/2$ denote the signal wavelength and antenna element spacing, respectively. Given the transmit signal model in \eqref{eq_passband_st}, the backscattered signal impinging onto the $\thnew{i}$ element of the radar RX array can be expressed as
\begin{align}\nonumber
    & y_i(t) = \alpha \left[  \arx(\theta) \right]_i \atx^T(\theta)  \ftx \sum_{m = 0}^{M-1} \sm\big(t - \tau(t)\big) e^{-j 2 \pi \fc \tau} e^{j 2 \pi \fc \nu t }~.
\end{align}

We make the following standard assumptions: \textit{(i)} the CP duration is larger than the round-trip delay of the furthermost target\footnote{We focus on small surveillance volumes where the targets are relatively close to the radar, such as vehicular applications.}, i.e., $\Tcp \geq \tau$, \cite{Firat_OFDM_2012,OFDM_Radar_Phd_2014,SPM_JRC_2019}, \textit{(ii)} the Doppler shifts satisfy $ \lvert \nu \rvert \ll 1/N$ \cite{Firat_OFDM_2012,ICI_OFDM_TSP_2020}, and \textit{(iii)} the time-bandwidth product (TBP) $B M \Tsym$ is sufficiently low so that the wideband effect can be ignored, i.e., $\sm(t - \tau(t)) \approx \sm(t - \tau)$ \cite{OFDM_ICI_TVT_2017}. Under this setting, sampling $y_i(t)$ at $t = m\Tsym + \Tcp + \ell T / N$ for $\ell = 0, \ldots, N-1$ (i.e., after CP removal for the $\thnew{m}$ symbol) and neglecting constant terms, the time-domain signal received by the $\thnew{i}$ antenna in the $\thnew{m}$ symbol can be written as \cite{ICI_OFDM_TSP_2020}
\begin{align}\label{eq_rec_bb2}
    y_{i,m}[\ell] &= \alpha \left[  \arx(\theta) \right]_i \atx^T(\theta) \ftx  \, e^{j 2 \pi \fc m \Tsym \nu  } e^{j 2 \pi \fc T \frac{\ell}{N} \nu} \\ \nonumber &~~\times \frac{1}{\sqrt{N}}  \sum_{n = 0}^{N-1}  \xnm \, e^{j 2 \pi n \frac{\ell}{N}} e^{-j 2 \pi n \deltaf \tau} ~.
\end{align}

\subsection{Fast-Time/Slow-Time Representation with ICI}
For the sake of convenience, let us define, respectively, the frequency-domain and temporal steering vectors and the ICI phase rotation matrix as
\begin{align} \label{eq_steer_delay}
	\bb(\tau) & \triangleq  \transpose{ \left[ 1, e^{-j 2 \pi \deltaf \tau}, \ldots,  e^{-j 2 \pi (N-1) \deltaf  \tau} \right] } \\ \label{eq_steer_doppler}
	\cc(\nu) & \triangleq \transpose{ \left[ 1, e^{-j 2 \pi f_c \Tsym \nu }, \ldots,  e^{-j 2 \pi f_c (M-1) \Tsym \nu } \right] } \\ \label{eq_ici_D}
	\DD(\nu) &\triangleq \diag{1, e^{j 2 \pi \fc \frac{T}{N} \nu}, \ldots, e^{j 2 \pi \fc \frac{T(N-1)}{N} \nu} } ~.
\end{align}
Accordingly, the radar observations in \eqref{eq_rec_bb2} can be expressed as
\begin{align} \label{eq_ym}
    \yy_{i,m} &= \alpha \, \left[  \arx(\theta) \right]_i \atx^T(\theta) \ftx  \DD(\nu) \FF_N^{H} \Big(\xx_m \odot \bb(\tau) \left[\cc^{*}(\nu)\right]_m  \Big)  
\end{align}
where $\FF_N \in \complexset{N}{N}$ is the unitary DFT matrix with $\left[ \FF \right]_{\ell,n} = \frac{1}{\sqrt{N}} e^{- j 2 \pi n \frac{\ell}{N}} $, $\yy_{i,m} \triangleq \left[ y_{i,m}[0] \, \ldots \, y_{i,m}[N-1] \right]^T$ and $\xx_m \triangleq \left[ x_{0,m} \, \ldots \, x_{N-1,m} \right]^T$.

\begin{figure}
	\centering
    \vspace{-0.2in}
	\includegraphics[width=0.7\linewidth]{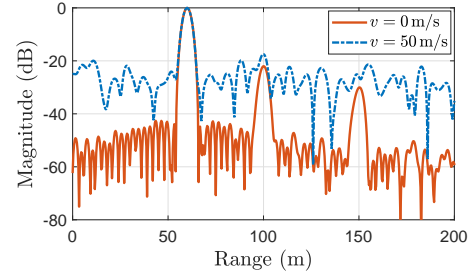}
	\vspace{-0.1in}
	\caption{\footnotesize Range profiles of MIMO-OFDM radar with the parameters given in Sec.~\ref{sec_sim} for two different target velocities. The scenario contains $3$ targets having the same velocity $v$, the ranges $(60, 100, 150) \, \rm{m}$, the angles $(25^\circ, 30^\circ, 35^\circ)$ and the SNRs (i.e., $\abs{\alpha_k}^2/\sigma^2$) $(30, 5, 0) \, \rm{dB}$, respectively. 
	}
	\label{fig_ici_comp_range_profile}
	\vspace{-0.22in}
\end{figure}

Aggregating \eqref{eq_ym} over $M$ symbols and considering the presence of multiple targets, the OFDM radar signal received by the $\thnew{i}$ antenna over a frame can be written in a fast-time/slow-time compact matrix form as
\begin{align} \label{eq_ym_all_multi}
    \boldY_i = \sum_{k=0}^{K-1} \alpha^{(i)}_k  \underbrace{\DD(\nu_k)}_{\substack{\rm{ICI} } } \FF_N^{H} \Big(\boldX \odot \bb(\tau_k) \cc^{H}(\nu_k) \Big)  + \boldZ_i
\end{align}
for $i = 0,\ldots, \Nrx-1$, where $\alpha^{(i)}_k \triangleq \alpha_k \, \left[  \arx(\theta_k) \right]_i \atx^T(\theta_k) \ftx$, $(\alpha_k, \tau_k, \nu_k, \theta_k)$ are the parameters of the $\thnew{k}$ target,
$\boldY_i \triangleq [ \yy_{i,0} \, \ldots \, \allowbreak \yy_{i,M-1} ] \in \complexset{N}{M}$, $\boldX \triangleq \left[ \xx_0 \, \ldots \, \xx_{M-1} \right] \in \complexset{N}{M}$ and $\boldZ_i \in \complexset{N}{M}$ is the additive noise matrix with $\vecc{\boldZ_i} \sim \mtCN(\boldzero_{NM}, \allowbreak \sigma^2 \Imatrix_{NM} ) $. In \eqref{eq_ym_all_multi}, each column contains fast-time samples within a particular symbol and each row contains slow-time samples at a particular range bin. The diagonal phase rotation matrix $\DD(\nu)$ quantifies the ICI effect in fast-time domain, leading to Doppler-dependent phase-shifts across fast-time samples of each OFDM symbol, similar to the CFO effect in OFDM communications \cite{Visa_CFO_TSP_2006,multiCFO_TSP_2019}. Fig.~\ref{fig_ici_comp_range_profile} illustrates the effect of ICI on the range profile of an OFDM radar.

Given the transmit data symbols $\boldX$, the problem of interest for OFDM radar sensing is to estimate channel gains $\{\alpha_k\}_{k=0}^{K-1}$, azimuth angles $\{\theta_k\}_{k=0}^{K-1}$, delays $\{\tau_k\}_{k=0}^{K-1}$ and Doppler shifts $\{\nu_k\}_{k=0}^{K-1}$ from the received $\Nrx \times N \times M$ space/fast-time/slow-time data cube $\{\boldY_i\}_{i=0}^{\Nrx-1}$ in \eqref{eq_ym_all_multi}.

\section{Parameter Estimation via APES Spatial Filtering}\vspace{-0.1in}
The ML approach to \eqref{eq_ym_all_multi} requires a computationally prohibitive $3K$ dimensional search over the parameter space. Additionally, the number of targets $K$ is unknown a-priori. To tackle this challenging estimation problem, we devise a three-stage low-complexity algorithm as described in the following subsections.
 
\subsection{Step 1: Angle Estimation via MUSIC}\vspace{-0.1in}
For mathematical convenience, we consider the space/fast-time snapshot of the data cube in \eqref{eq_ym_all_multi} corresponding to the $\thnew{m}$ OFDM symbol:
\begin{align}\label{eq_ybbar}
    \Ybbar_m &\triangleq \left[ \yy_{0,m} \, \ldots \, \yy_{\Nrx-1,m}  \right] \in \complexset{N}{\Nrx} \\ \nonumber
    &= \sum_{k=0}^{K-1} \alpha_k \,  \atx^T(\theta_k) \ftx \DD(\nu_k) \FF_N^{H} \diag{\xx_m} \\ \nonumber &~~~~~~~~\times \bb(\tau_k) \left[ \cc^{*}(\nu_k) \right]_m \arx^T(\theta_k)  + \Zbbar_{m}
\end{align}
for $m=0,\ldots,M-1$. We propose to perform angle estimation using the MUSIC algorithm \cite{MUSIC_1986}. To this end, we first construct the spatial covariance matrix (SCM) of the data cube in \eqref{eq_ybbar} as
\vspace{-0.05in}
\begin{equation}\label{eq_SCM_obs}
    \boldR \triangleq \sum_{m=0}^{M-1} \Ybbar_m^H \Ybbar_m ~.
    \vspace{-0.05in}
\end{equation} 
Under the assumption of spatially non-overlapping targets (i.e., $\arx^H(\theta_{k_1}) \arx(\theta_{k_2}) \approx 0, \, \forall \, k_1 \neq k_2$), the SCM can be approximated as (the proof is omitted due to space limitation)
\vspace{-0.05in}
\begin{equation}
    \boldR \approx \norm{\boldX}_F^2 \sum_{k=0}^{K-1} \abs{\alpha_k}^2 \abs{\atx^T(\theta_k)\ftx}^2 \arx^{*}(\theta_k)  \arx^T(\theta_k) + \sigma^2 \Imatrix  
    \vspace{-0.05in}
\end{equation}
which is independent of target delays and Dopplers. Hence, creating the MUSIC spectrum from the SCM in \eqref{eq_SCM_obs} does not require target delay-Doppler information. Assuming $\Nrx > K$, let the eigendecomposition of the SCM be denoted as $\boldR = \boldU_s \boldLambda_s \boldU^H_s + \boldU_n \boldLambda_n \boldU^H_n$, where the diagonal matrix $\boldLambda_s$ contains the $K$ largest eigenvalues, $\boldLambda_n$ contains the remaining $\Nrx-K$ eigenvalues, and $\boldU_s$ and $\boldU_n$ have the corresponding eigenvectors as their columns. Then, the MUSIC spectrum can be computed as 
\vspace{-0.05in}
\begin{align} \label{eq_spatial_spectrum}
    f(\theta) &= \frac{ 1  }{   \arx^T(\theta) \boldU_n \boldU_n^H  \arx^{*}(\theta) } ~.
    \vspace{-0.05in}
\end{align}
Let $\Smcal = \{ \thetahat_0, \ldots, \thetahat_{K-1} \}$ be the set of estimated angles in Step~1, which correspond to the peaks of the MUSIC spectrum\footnote{To prevent performance degradation at low SNRs due to spurious peaks and misidentification of signal and noise subspaces, improved versions of MUSIC can be employed, e.g., \cite{SSMUSIC_2002}.} in \eqref{eq_spatial_spectrum}.

\subsection{Step 2: Joint Doppler/CFO and Unstructured Channel Estimation via APES Beamforming}\vspace{-0.1in}
In Step~2, we formulate a joint Doppler/CFO and channel estimation problem for each $\thetahat \in \Smcal$, assuming the existence of a single target at a given angle. Invoking the assumption of spatially non-overlapping targets, we treat interferences from other target components as noise and consider a single-target model in \eqref{eq_ybbar} for each $\thetahat \in \Smcal$. To that aim, let
\begin{equation}\label{eq_H_dec}
    \boldH = \left[ \hh_0 \, \ldots \, \hh_{M-1} \right]  \in \complexset{L}{M}
\end{equation}
denote the unstructured, single-target radar channels in the time domain with $L$ taps, collected over $M$ OFDM symbols. Here, $L \leq N \Tcp / T$ due to the CP requirement. Based on this unstructured representation, \eqref{eq_ybbar} can be re-written as
\vspace{-0.05in}
\begin{align} \label{eq_ym_all_single2}
    \Ybbar_m =  \DD(\nu) \Xbbar_m \hh_m  \arx^T(\thetahat)  + \Zbbar_{m} 
    \vspace{-0.05in}
\end{align}
where $\Xbbar_m \triangleq \FF_N^{H} \diag{\xx_m} \FF_{N,L}$, $\FF_{N,L} \in \complexset{N}{L}$ denotes the first $L$ columns of $\FF_N$ and $\Zbbar_{m}$ contains noise and interferences from other targets in $\Smcal$. According to \eqref{eq_ybbar}, the frequency-domain radar channels have the form
\begin{equation}\label{eq_H_def}
    \FF_{N,L} \boldH = \alphabar \, \bb(\tau) \cc^{H}(\nu) 
\end{equation}
with $\alphabar \triangleq \alpha \, \atx^T(\thetahat) \ftx$ representing the complex channel gain including the transmit beamforming effect. \vspace{-0.1in}

\begin{remark}[Duality Between OFDM Communications and OFDM Radar in \eqref{eq_ym_all_single2}]\label{remark_down_comm}
Based on the observation that radar targets can be interpreted as uncooperative users from a communications perspective (as they transmit information to the radar receiver via reflections in an unintentional manner \cite{jointRadCom_review_TCOM,chiriyath2017radar}), we point out an interesting duality between the OFDM radar signal model with ICI in \eqref{eq_ym_all_single2} and an OFDM communications model with CFO (e.g., \cite[Eq.~(5)]{multiCFO_TWC_2018} and \cite[Eq.~(4)]{zhang2014blind}). Precisely, $\DD(\nu)$ represents CFO between the OFDM transmitter and receiver for a communications setup, while it quantifies the ICI effect due to high-speed targets for OFDM radar. Similarly, $\Xbbar_m$ represents data/pilot symbols for communications and probing signals for radar\footnote{For radar sensing, every symbol acts as a pilot due to dual-functional operation on a single hardware platform.}. In addition, $\hh_m$ represents the time-domain channel for communications and the structured (delay-Doppler parameterized) channel for radar. 
\end{remark}
\vspace{-0.1in}
In light of Remark~\ref{remark_down_comm}, we re-formulate the radar delay-Doppler estimation problem as a communication channel estimation problem, where the objective is to jointly estimate the unstructured time-domain channels $\boldH$ and the CFO $\nu$ from \eqref{eq_ym_all_single2}. To perform channel estimation in \eqref{eq_ym_all_single2}, we propose an APES-like beamformer \cite{Est_MIMO_radar_2008}
\begin{align} \label{eq_apes}
\mathop{\mathrm{min}}\limits_{\ww, \boldH, \nu} &~~ \sum_{m=0}^{M-1}
\normbig{ \Ybbar_m \wwconj -  \DD(\nu) \Xbbar_m \hh_m  }^2  \\ \nonumber
\mathrm{s.t.}&~~ \ww^H \arx(\thetahat) = 1 
\end{align}
where $\ww \in \complexset{\Nrx}{1}$ is the APES spatial beamforming vector for an estimated angle $\thetahat \in \Smcal$. The optimal channel estimate for the $\thnew{m}$ symbol in \eqref{eq_apes} for a given $\ww$ and $\nu$ is given by
\begin{equation}\label{eq_hm_est}
    \hhhat_m = \Big( \Xbbar_m^H \Xbbar_m \Big)^{-1} \Xbbar_m^H \DD^H(\nu) \Ybbar_m \wwconj ~.
\end{equation}
Plugging \eqref{eq_hm_est} back into \eqref{eq_apes} yields
\begin{align} \label{eq_apes2}
\mathop{\mathrm{min}}\limits_{\ww, \nu} &~~ \ww^T \boldQ(\nu) \wwconj ~~~~~~ \mathrm{s.t.}~~ \ww^H \arx(\thetahat) = 1 
\end{align}
where
    $\boldQ(\nu) \triangleq \boldR - \boldSigma(\nu)$
is the residual SCM, $\boldR$ is the SCM of the observed data cube in \eqref{eq_SCM_obs} and
\begin{equation}\label{eq_cfo_covariance}
    \boldSigma(\nu) \triangleq \sum_{m=0}^{M-1} \Big( \projrange{\Xbbar_m} \DD^H(\nu) \Ybbar_m \Big)^H \Big( \projrange{\Xbbar_m} \DD^H(\nu) \Ybbar_m \Big)
\end{equation}
is the SCM of the CFO compensated observed data component that lies in the subspace spanned by the columns of the pilot symbols. 
For a given CFO $\nu$, the optimal beamformer in \eqref{eq_apes2} can be obtained in closed-form as \cite{Est_MIMO_radar_2008}
\begin{equation}\label{eq_what}
    \wwhat = \frac{ \boldQ^{*}(\nu)^{-1} \arx(\thetahat) }{ \arx^H(\thetahat) \boldQ^{*}(\nu)^{-1} \arx(\thetahat)  }~.
\end{equation}
Substituting \eqref{eq_what} into \eqref{eq_apes2}, the CFO can be estimated as
\begin{equation}\label{eq_nuhat}
    \nuhat = \arg \max_{\nu} ~~ \arx^H(\thetahat) \boldQ^{*}(\nu)^{-1} \arx(\thetahat) ~.
\end{equation}
Finally, plugging \eqref{eq_what} and \eqref{eq_nuhat} into \eqref{eq_hm_est}, the channel estimates can be expressed as
\begin{equation}\label{eq_hm_est2}
    \hhhat_m = \frac{ \Big( \Xbbar_m^H \Xbbar_m \Big)^{-1} \Xbbar_m^H \DD^H(\nuhat) \Ybbar_m  \boldQ(\nuhat)^{-1} \arx^{*}(\thetahat) }{ \arx^T(\thetahat) \boldQ(\nuhat)^{-1} \arx^{*}(\thetahat)   }~.
\end{equation}

\subsection{Step 3: Delay-Doppler Recovery from Unstructured Channel Estimates}\vspace{-0.1in}
Given the unstructured channel estimates $\boldHhat \triangleq \left[ \hhhat_0, \ldots, \hhhat_{M-1} \right]$ obtained in \eqref{eq_hm_est2}, we aim to estimate channel gain, delay and Doppler shift via a least-squares (LS) approach by exploiting the structure in \eqref{eq_H_def} as follows:
\begin{align} \label{eq_apes_step2}
\mathop{\mathrm{min}}\limits_{\alpha, \tau, \nu} &~~ 
\normbig{ \FF_{N,L} \boldHhat -  \alphabar \, \bb(\tau) \cc^H(\nu)  }_F^2 ~.
\end{align}
In \eqref{eq_apes_step2}, delay and Doppler estimates $\tauhat$ and $\nuhat$ can be obtained simply via 2-D FFT (i.e., IFFT and FFT across the columns and rows of $\FF_{N,L}  \boldHhat$, respectively). Then, channel gain can be estimated as $\alphabarhat =  \bb^H(\tauhat) \FF_{N,L} \boldHhat  \cc(\nuhat)  /  (\norm{\bb(\tauhat)}^2 \norm{\cc(\nuhat)}^2 )$. The overall algorithm is summarized in Algorithm~\ref{alg_apes}.

\begin{algorithm}
	\caption{APES Filtering for ICI-Aware MIMO-OFDM Radar}
	\label{alg_apes}
	\begin{algorithmic}
	    \State \textbf{Input:} Space/fast-time/slow-time data cube $\{\boldY_i\}_{i=0}^{\Nrx-1}$ in \eqref{eq_ym_all_multi}.
	    \State \textbf{Output:} Delay-Doppler-angle triplets $\{ (\tauhat_k, \nuhat_k, \thetahat_k) \}_{k=0}^{K-1}$.
	    \State \textbf{Step~1:} Estimate target angles by identifying the peaks in the MUSIC spatial spectrum in \eqref{eq_spatial_spectrum}.
	    \State \textbf{Step~2:} For each estimated angle $\thetahat$: 
	    \State \hskip1.0em Estimate the Doppler/CFO from \eqref{eq_nuhat}. 
	    \State \hskip1.0em Estimate the time-domain channels $\boldHhat$ via \eqref{eq_hm_est2}.
	    \State \textbf{Step~3:} For each estimated angle $\thetahat$, estimate delay-Doppler in \eqref{eq_apes_step2} from the unstructured channel estimates $\boldHhat$.
	\end{algorithmic}
	\normalsize
\end{algorithm}

\vspace{-0.2in}
\section{Simulation Results}\label{sec_sim}
\begin{figure}
	\centering
    \vspace{-0.2in}
	\includegraphics[width=0.8\linewidth]{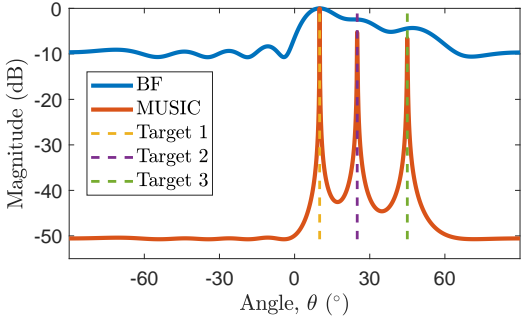}
	\vspace{-0.1in}
	\caption{\footnotesize MUSIC spatial spectrum of OFDM radar in Step~1 along with the results of ordinary beamforming (BF) $\arx^T(\theta) \boldR \arx^{*}(\theta)$.}
	\label{fig_music_step1}
	\vspace{-0.15in}
\end{figure}

\begin{figure}
	\centering
    \vspace{-0.15in}
	\includegraphics[width=1.1\linewidth]{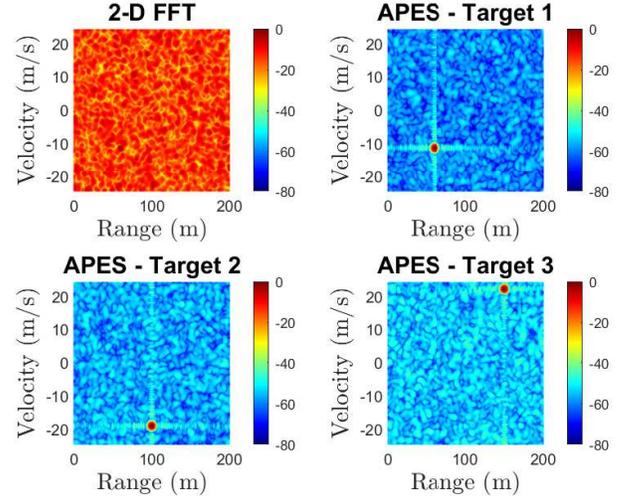}
	\vspace{-0.3in}
	\caption{\footnotesize Range-velocity profiles obtained by standard 2-D FFT after receive beamforming towards $10^\circ$ and those obtained by Algorithm~\ref{alg_apes} in Step~3 as the output of 2-D FFT of target-specific frequency-domain channel estimates for Target~1, Target~2 and Target~3.}
	\label{fig_range_vel_step3}
	\vspace{-0.1in}
\end{figure}

\vspace{-0.1in}
To demonstrate the performance of Algorithm~\ref{alg_apes}, we consider an OFDM system with $\fc = 60 \, \rm{GHz}$, $B = 50 \, \rm{MHz}$, $N = 2048$, $\deltaf = 24.41 \, \rm{kHz}$, $\Tsym = 51.2 \, \rm{\mu s}$, $M = 64$, $\Ntx = 8$ and $\Nrx = 8$.
The data symbols $\boldX$ are randomly generated from the QPSK alphabet and the transmit beamformer is set to point towards $30^\circ$, i.e., $\ftx = \atx^{*}(30^\circ)$. 
To illustrate the output of Algorithm~\ref{alg_apes}, we first consider a scenario consisting of $K=3$ targets with the ranges $(60, 100, 150) \, \rm{m}$, the velocities $(-60, 30, 120) \, \rm{m/s}$, the angles $(10^\circ, 25^\circ, 45^\circ)$ and the SNRs (i.e., $\abs{\alpha_k}^2/\sigma^2$) $(30, 15, 25) \, \rm{dB}$. Fig.~\ref{fig_music_step1} shows the MUSIC spectrum \eqref{eq_spatial_spectrum} obtained in Step~1 of Algorithm~\ref{alg_apes}. 
In addition, Fig.~\ref{fig_range_vel_step3} demonstrates the range-velocity profiles obtained in Step~3 for each target angle along with the results of standard 2-D FFT \cite{RadCom_Proc_IEEE_2011}. It is observed that the proposed algorithm can successfully separate multiple target reflections in the angular domain, estimate their Dopplers/CFOs for ICI elimination and accurately recover delays and Dopplers.

\begin{figure}
	\centering
	\includegraphics[width=0.82\linewidth]{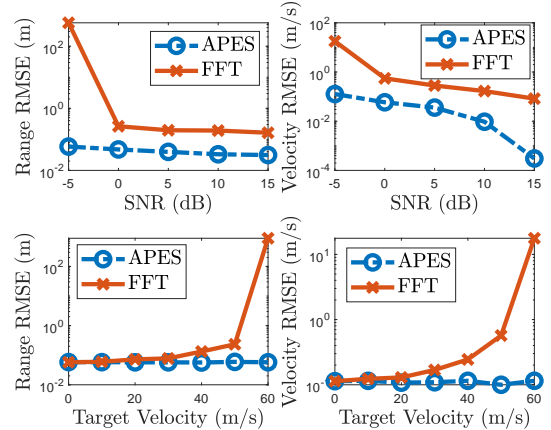} \vspace{-0.15in}
	\caption{Range and velocity estimation performances of Algorithm~1 and 2-D FFT benchmark.}
	\label{fig_single_target_comparison}
	\vspace{-0.1in}
\end{figure}

Second, we compare the performance of Algorithm~\ref{alg_apes} with the 2-D FFT benchmark \cite{RadCom_Proc_IEEE_2011} in a single-target scenario with $(R, v, \theta) = (80\, \rm{m}, 70\, \rm{m/s}, 30^\circ)$ and an SNR of $-5 \, \rm{dB}$ using $M=8$ symbols. Since there exists no previous estimators for ICI-aware MIMO-OFDM radar, 2-D FFT is applied on the fast-time/slow-time snapshot obtained by receive beamforming of the data cube towards the true target angle. Fig.~\ref{fig_single_target_comparison} shows the range and velocity RMSEs of Algorithm~\ref{alg_apes} and the 2-D FFT benchmark with respect to SNR and target velocity over $100$ Monte Carlo noise realizations. As seen from the figure, ICI-induced high side-lobe levels in the delay-Doppler domain significantly degrade the performance of the 2-D FFT algorithm, while the proposed algorithm can suppress the ICI effect by exploiting the signal structure within an APES beamforming framework.



\vspace{-0.15in}
\section{Conclusion}\vspace{-0.15in}
This paper addresses the parameter estimation problem for a MIMO-OFDM radar in the presence of non-negligible ICI caused by high-mobility targets. Based on an APES spatial filtering approach, a novel delay-Doppler-angle estimation algorithm is proposed by re-formulating radar sensing as a communication channel estimation problem. Simulation results show that the proposed algorithm enables high-accuracy multi-target parameter estimation under strong ICI by separating individual target reflections in the angular domain.




\vfill\pagebreak



\bibliographystyle{IEEEbib}
\bibliography{main}

\end{document}